# X-ray microtomography in biology

Ryuta Mizutani [a,] *, Yoshio Suzuki [b]

[a] Department of Applied Biochemistry, School of Engineering, Tokai University, Kitakaname 4-1-1, Hiratsuka, Kanagawa 259-1292, Japan
[b] Research and Utilization Division, JASRI/SPring-8, Kouto 1-1-1, Sayo, Hyogo 679-5198, Japan

*Corresponding author. Tel: +81-463-58-1211; fax: +81-463-50-2506.
E-mail address: ryuta@tokai-u.jp (R. Mizutani).

**Abstract**

Progress in high-resolution x-ray microtomography has provided us with a practical approach to determining three-dimensional (3D) structures of opaque samples at micrometer to submicrometer resolution. In this review, we give an introduction to hard x-ray microtomography and its application to the visualization of 3D structures of biological soft tissues. Practical aspects of sample preparation, handling, data collection, 3D reconstruction, and structure analysis are described. Furthermore, different sample contrasting methods are approached in detail. Examples of microtomographic studies are overviewed to present an outline of biological applications of x-ray microtomography. We also provide perspectives of biological microtomography as the convergence of sciences in x-ray optics, biology, and structural analysis.

**1. Introduction**

Progress in high-resolution x-ray microtomography (also known as microcomputed tomography or micro-CT) has provided us with a practical approach to determining the three-dimensional (3D) structure of an opaque sample at micrometer to submicrometer resolution (Bonse and Busch, 1996; Salomé et al., 1999; Uesugi et al., 2001; Takeuchi et al., 2002). Applications of x-ray microtomography have been reported for a wide variety of objects (Toda et al., 2008; Chen et al., 2009; Fusseis et al., 2009; Tsuchiyama et al., 2011; Zhu et al., 2011). In biology, x-ray microtomography has revealed 3D structures of biological samples from many species of organisms including human (Bonse et al., 1994; Salomé et al., 1999; Mizutani et al., 2008a), mouse (Johnson et al., 2006; de Crespigny et al., 2008; Mizutani et al., 2009a), and insect (Mizutani et al., 2007; Metscher, 2009; van de Kamp et al., 2011). It has become a common method in studies associated with osteo and dental microstructures (Neues and Epple, 2008; Zou et al., 2011). Microtomographic studies of soft tissues, which account for a major proportion of biological tissues, have shed light on the structural mechanism of biological functions (Happel et al., 2010; Mizutani et al., 2010a).

The constituents of soft tissue are cells and extracellular matrices that are responsible for biological functions. For example, the cellular structures of brain tissue or subcellular structures of nucleus are essential for their activities. In this review, we focus on hard x-ray microtomography of biological soft tissues at cellular to subcellular resolution, i.e., micrometer to submicrometer resolution. The subcellular and cellular microstructures build up a 3D structure, called tissue, with dimensions of the order of hundreds of micrometers.

Although microscopy using visible light is the primary method for visualizing structures of biological systems, absorbance and refraction of the visible light interferes with the determination of the internal structures of real biological objects and this interference increases with tissue thickness. Thus, light microscopy is mainly used for imaging sectioned samples. By contrast, the transmissive and less refractile nature of hard x-rays with respect to biological tissue enables radiographic observation of the 3D structure. This allows microtomographic studies of the 3D structure of biological objects that have not been visualized before.

Below, we discuss the fundamentals of x-ray microtomography in biology for those who are not familiar with microtomography. Practical aspects of biological sample preparation, handling, data collection, 3D reconstruction, and structure analysis are described. Examples of microtomographic studies are overviewed to present an outline of biological applications of x-ray microtomography. In the last section, we give perspectives of x-ray microtomography in biology as the convergence of sciences in x-ray optics, biology, and structural analysis.

**2. Microtomography in practice**

*2.1. X-ray visualization of biological tissue*

Since x-rays interact with electrons in sample objects, electron-dense or electron-poor structures can be visualized in an x-ray image as contrast against the background. However, biological soft tissue is composed of low-atomic-number (low-Z) elements, such as carbon and oxygen, which produce little contrast in a hard x-ray image. Most biological soft tissues exhibit uniform density in a conventional x-ray image in which only the outline but no internal constituents are visualized. Although the structural outline is biologically relevant in some cases, the internal structure is crucial for the biological function of soft tissue.

In order to visualize soft tissue effectively, we should label the structure of interest with a probe appropriate for the observation method. In modern biological studies using light microscopy, the standard procedure for visualizing the target structure involves labeling the sample with a visible dye or fluorescent probe. Similarly, x-ray-specific labeling should be performed for x-ray visualization except in circumstances in which sample pretreatment is not feasible such as in the case of fossils (Friis et al., 2007). In clinical diagnosis, a possible visceral lesion behind the skin is examined using x-rays. Although soft tissues of the human body give little contrast in an x-ray image, luminal structures can easily be visualized by using x-ray contrast media. These contrast media contain high-atomic-number (high-Z) elements, such as iodine or barium, which absorb x-rays efficiently. In the same way, x-ray visualization of the microstructures of soft tissues can be performed by specifically labeling each biological constituent with a high-Z element probe. A number of labeling methods in biological microtomography have been reported, as overviewed in the following section.

In x-ray microtomography, the interaction of x-rays with a sample object is mostly visualized by one of three types of modality: absorption, phase interference, or fluorescence. Which of these modalities should be used for visualizing samples labeled with high-Z element probes? The x-ray fluorescence of a probe element down to the concentration of 1 ppm can be visualized three-dimensionally (Lanzirotti et al., 2010), although it takes much more time to acquire a fluorescence image compared with the usual x-ray image. Hence, the application of fluorescence microtomography is still limited to a small number of examples at present. Microtomographic studies using phase contrast methods have been reported for a number of biological samples (Betz et al., 2007; Connor et al., 2009; Wu et al., 2009; Schulz et al., 2010). However, the prerequisite for these applications is that the sample objects consist only of low-Z elements, i.e., without any high-Z element probes. On the other hand, a structure labeled with a high-Z element probe is rather effectively visualized in an absorption contrast image in comparison with a phase contrast image. This is because the x-ray absorption is approximately proportional to the product of electron density and the cube of atomic number $Z$, while the x-ray phase shift is approximately proportional to only the electron density. From these considerations, we recommend absorption contrast microtomography as the primary method for visualizing high-Z element probes.

*2.2. High-Z element probes*

Element probes that have been reported for visualizing biological microstructures include osmium (Ananda et al., 2006; Johnson et al., 2006; Litzlbauer et al., 2006; Lareida et al., 2009; Kamenz and Weidemann, 2009; Mizutani et al., 2009b), gold (Mizutani et al., 2006, 2007, 2008b, 2008c, 2009b, 2010b; Li et al., 2010; Nelson et al., 2010; Wang et al., 2011), silver (Mizutani et al., 2008a, 2008b, 2010a; Parameswaran et al., 2009), iodine (de Crespigny et al., 2008; Metscher, 2009; Lee et al., 2010; Jeffery et al., 2011), platinum (Mizutani et al., 2008b, 2008c), mercury (Mizutani et al., 2009b), tungsten (Metcher, 2009), and lead (Kamenz and Weidemann, 2009). These elements, which are in the fifth or sixth row of the periodic table, can give sufficient contrast in an x-ray image represented with linear absorption coefficients (LACs). Elements with atomic numbers of 67 to 83, including osmium, gold, platinum, mercury, tungsten, and lead, have their $L_{III}$ absorption edges between 8 keV and 14 keV, which falls within the range of hard x-ray energy typically available at synchrotron radiation facilities. In this section, we discuss the characteristics of each probe element and give a labeling method.

*2.2.1. Osmium*

Osmification using osmium tetroxide has been applied to microtomographic visualization of soft tissues (Ananda et al., 2006; Johnson et al., 2006; Litzlbauer et al., 2006; Lareida et al., 2009; Kamenz and Weidemann, 2009; Mizutani et al., 2009b). It is a conventional fixation and staining method in biological electron microscopy. Its characteristics are permeability and uniformity. Osmified tissue gives uniformly stained images even for block samples with millimeter-scale thickness. Since plasma membrane lipids are supposedly labeled with osmium tetroxide, osmification should visualize all cells. Consequently, it is difficult to distinguish tissue structure constituents from each other compared with other labeling methods. Fig. 1 shows an example of a brain tissue microstructure visualized by osmification (Mizutani et al., 2009b). In this figure, fibriform structures that are slightly denser than the surroundings are visualized.

*Method for osmification.* Incubate the sample overnight at room temperature while rocking it in a solution containing 0.1 M sodium cacodylate (pH 7.2), 1% glutaraldehyde, and 1% osmium tetroxide. Osmium tetroxide requires careful handling according to the CDC guidelines (http://www.cdc.gov/niosh/topics/chemical.html). Wash the sample for 30 min in 0.1 M sodium cacodylate buffer and wash twice more for 30 min each time in phosphate-buffered saline. Then dehydrate the sample in an ethanol series (Johnson et al., 2006).

*2.2.2. Gold*

Gold and silver impregnation is a conventional procedure in light microscopy observation of soft tissues (Grizzle, 1996). The first step is silver deposition in which argyrophil constituents are labeled with silver ions. The adsorbed silver is then substituted with gold metal particles, which can be visualized in the x-ray image (Mizutani et al., 2006, 2007). One of the gold impregnation methods used in x-ray microtomography is reduced-silver Bodian impregnation, which visualizes every neuron by staining neuropils. A microtomogram of human frontal cortex stained with this method (Mizutani et al., 2010b) indicates that neuropil networks and small spherical structures in the cellular nucleus can be visualized. A number of variations of the gold and silver impregnation can be applied to microtomographic visualization. Reticular fiber structures of connective tissue can be visualized by reticulin silver impregnation (Mizutani et al., 2009b). Fig. 2 shows another example of a gold-labeled structure of intestinal villi of adult C57BL/6 mouse visualized by the periodic acid-methenamine-silver stain method.

Gold impregnation has been developed for staining sectioned samples. In some cases, the permeation of gold is limited to a depth of approximately 100 μm from the tissue surface. This permeation limit can be overcome by using detergents with aurate reagent, by defatting sample tissue prior to staining, or by using tissues from juvenile animals.

Colloidal gold has also been reported as a high-Z element probe for x-ray microtomography. Gold-antibody conjugates have been used as electron-dense markers in electron microscopy (Mayhew et al., 2009). Colloid itself shows adsorption on neural tissue and has been visualized by x-ray microtomography (Mizutani et al., 2008c). Selective visualization of biological constituents with gold conjugates has also been reported (Mizutani et al., 2009b; Li et al., 2010; Nelson et al., 2010; Wang et al., 2011).

*Method for Bodian impregnation.* Wash fixed tissue in distilled water. Incubate the tissue for 15 hours at 37°C in a solution containing 0.04% silver nitrate, 0.75% pyridine, and 25 mM sodium borate (pH 8.4). Develop the deposited silver for 10 min at 25°C in a solution containing 1% hydroquinone and 10% sodium sulfite. After it has been washed with distilled water twice, immerse the tissue sequentially in 1% hydrogen tetrachloroaurate for 60 min, 2% oxalic acid for 10 min, and 5% sodium thiosulfate for 5 min, with washes in distilled water between immersions. After removal of the residual sodium thiosulfate by washing, dehydrate the tissue in an ethanol series (Mizutani et al., 2007).

*2.2.3. Silver*

Though silver is in the 5th row of the periodic table and exhibits a lower LAC (776 $cm^{-1}$ at 12 keV) than 6th-row elements (e.g., 3470 $cm^{-1}$ at 12 keV for gold), high-density deposition of silver compound in soft tissue can be visualized in a microtomogram.

Golgi silver impregnation (Valverde, 1970) is a conventional method for light microscopy observation of neurons. While this method uses dichromate and silver ions as labeling probes, silver should mainly contribute to the x-ray absorption. Microtomographic analysis of the 3D structure of human cerebral cortex stained with this method has been reported (Mizutani et al., 2008a). Fig. 3 shows an example of neuron structures visualized by Golgi impregnation. Neuronal circuits of human brain tissue were determined from the obtained structure (Mizutani et al., 2010a). This strategy using high-Z element labeling and subsequent microtomographic analysis is the only method of determining human cerebral circuits composed of neurons. Although Golgi impregnation can stain the internal structure of block tissue, only a limited population of its constituents (approximately 10% of neurons) is visualized (Mizutani et al., 2010a). A selective staining mechanism has not been revealed clearly.

Silver staining commonly used for staining proteins in polyacrylamide gels has been used to visualize small alveoli and thin septal walls of mouse lung (Parameswaran et al., 2009). Microtomographic visualization of the lung tissue labeled with silver allowed the quantification of alveolar airspace volume.

*Method for Golgi impregnation.* Dissect tissue sample into a 10-mm block. Wash the tissue for 5 min in a solution containing 2.5% potassium dichromate and 4% formaldehyde and then incubate for 7 days at 25ºC in a solution containing 2.5% potassium dichromate. After blotting with filter paper and a brief (a few seconds) wash in distilled water, incubate the tissue for 48 hours at 25ºC in 0.75% silver nitrate. Wash residual silver nitrate with distilled water. Repeat these potassium dichromate and silver nitrate treatment steps three times. Then dehydrate the tissue in an ethanol series (Mizutani et al., 2010a).

*2.2.4. Iodine*

Soft tissues treated with iodine reagents can be visualized by x-ray microtomography. Elemental iodine permeates into soft tissue and adsorbs uniformly to the tissue constituents. The resultant image gives the overall structure of the sample. Vertebrate and invertebrate structures have been reported by using this method (Metcher, 2009; Jeffery et al., 2011). An iodinated contrast reagent 3,5-diacetamido-2,4,6-triiodobenzoic acid called Hypaque, has been used for visualizing mouse and rabbit brains (de Crespigny et al., 2008). Another contrast reagent, iopamidol encapsulated into poly(vinyl alcohol) microparticles, has also been reported as an iodine probe to visualize biofluid flows (Lee et al., 2010).

*Method for iodine staining.* Wash fixed tissue in distilled water. Incubate the tissue in aqueous solution containing 0.1% iodine and 0.2% potassium iodide overnight. Wash the stained tissue with water and perform microtomographic data collection while immersed in water (Metcher,

2009).

*2.2.5. Other elements*

Several other elements have been reported for microtomographic visualization of biological soft tissues. Platinum can be used as a substitute for gold. Element-specific visualization has been performed by double staining with gold and platinum (Mizutani et al., 2008b, 2008c). Phosphotungstic acid has been used with hematoxylin dye to stain sectioned tissue in light microscopy. This method can visualize the overall structure since phosphotungstic acid binds to various proteins. It has visualized the structures of vertebrate and invertebrate samples (Metcher, 2009). Biofilm in porous media has been distinctively visualized by using potassium iodide and barium sulfate suspension (Davit et al., 2011). Lead compounds exhibit adsorption on insect tissue (Kamenz and Weidemann, 2009). B-5 fixative containing mercuric chloride can be used for labeling cerebral cortex tissue (Mizutani et al., 2009b). Uranyl acetate has been used in combination with other reagents (Ananda et al., 2006; Faraj et al., 2009). Titan nanoparticles conjugated with doxorubicin have been used in x-ray fluorescence mapping of the particle localization in a carcinoma cell (Arora et al., 2010). Iron oxide nanoparticles have been used for microtomographic visualization of cardiac progenitor cell homing in infarcted rat hearts (Giuliani et al., 2011).

*2.2.6. Required concentration of probe elements in soft tissue*

Structural constituents can be recognized if the structure of interest has higher contrast than the image noise. Since an absorption contrast image taken using monochromatic x-rays is expressed in LACs, each voxel of the image represents the LAC of the corresponding position in the sample. Noise in the absorption contrast image is given by the deviation of the LAC. Fig. 4 shows an example of an absorption contrast microtomogram taken at the BL20XU beamline of the synchrotron radiation facility SPring-8 (Hyogo, Japan) using 12-keV x-rays. This example shows a cross-section of homogeneous epoxy resin filling a borosilicate glass capillary. The voxel dimension of this microtomogram is 0.500 μm in each direction. LACs of a 100 μm × 100 μm region of the resin indicated with the box in this figure gave an average LAC of 1.1 cm$^{-1}$ and deviation of 1.3 cm$^{-1}$. Therefore, LAC differences higher than approximately 1.3 cm$^{-1}$, if any exist, can be recognized as structures in this microtomogram.

Next, let us consider the effect of the voxel size with respect to the image's noise level. If we take samples $X_i$ of size $n$ from a certain population with variance $\sigma^2$, the average of the sample

$$\overline{X} = \frac{1}{n}\sum_{i=1}^{n} X_i$$

gives the variance of the average $\overline{X}$ as

$$\sigma^2(\overline{X}) = \frac{\sigma^2}{n}.$$

Digitizing images into voxels is equivalent to taking samples within the voxel from the presampled native image and averaging them within the voxel. In this digitizing process, the sample size taken from the native image population is represented by the voxel volume. Therefore, assuming the same brilliance and fluctuation for the incident x-rays, the same signal-to-noise ratio of the detector, and the same exposure duration, the variance of a microtomogram with isotropic voxel dimension $w$ μm can be written as

$$\sigma^2\langle w\rangle = \frac{\sigma_0^2}{w^3}$$

where $\sigma_0$ represents the deviation of voxel values in the microtomogram for a voxel volume of 1 μm$^3$. Hence, the deviation of voxel values, i.e., noise level of the microtomogram, can be expressed as

$$\sigma\langle w\rangle = \frac{\sigma_0}{w^{3/2}}.$$

This indicates that the noise level of the microtomogram is inversely proportional to the voxel dimension raised to the power of 3/2. In our experiments using 12-keV x-rays at the BL20XU beamline of SPring-8, a microtomogram with a voxel dimension of 0.5 μm and total exposure of 500 s typically gave an approximate voxel deviation of 1 cm$^{-1}$.

The degree of staining with the high-Z element should be determined on the basis of these considerations. Gold, which is our first choice for the probe element, exhibits an LAC of 3470 cm$^{-1}$ at x-ray energy of 12 keV. If 0.1% of the voxel volume is occupied with metal gold, the voxel value falls to 3.5 cm$^{-1}$, approximately three times the typical voxel value deviation. In this case, the stained structure can be clearly distinguished from the background. Gold concentration lower than 0.03%(v/v) would result in an ambiguous image that could not be recognized as a structure. Over-staining would give rise to difficulties in detecting transmitted x-rays through the stained structure. Therefore, probe concentration estimation and its feedback to the staining procedure as well as the selection of an effective probe are important for appropriate visualization of the structure of interest. The average filling occupancy of metal gold has been reported to be 0.24%(v/v) in nerve tissue visualized by using Bodian impregnation (Mizutani et al., 2007).

*2.3. Alternatives to high-Z element labeling*

Some biological tissues exhibit inherent contrast in an x-ray image. A number of microtomographic studies on bones, teeth, and mineralized tissues composed of calcium compounds have been reported (Neues and Epple, 2008; Zou et al., 2011). X-ray absorption by fat is lower than that by protein, giving the 3D distribution of fat in muscle tissues (Frisullo et al., 2010). X-ray microtomography of human lung tissue has visualized peripheral airways and airspaces that exhibit lower densities than alveolar walls do (Ikura et al., 2004).

Another method of visualizing biological structures is critical point drying, which has been used for preparing electron microscopy samples. In this method, volatile components such as water contained in the sample are replaced with air voids, giving an x-ray image of the residual constituents. Since changes in sample size have been observed for critical-point-dried samples (Kääb et al., 1998), the boundary of non-volatile segments would be visualized by the sample shrinkage. Although a number of microtomographic studies of critical-point-dried samples have been reported (Happel et al., 2010; van de Kamp et al., 2011), it should be noted that structural distortions can be introduced by the critical-point-drying procedure (e.g., Small et al., 2008).

Corrosion casting with resin has been used for visualizing the 3D microstructure of vascular systems (Mondy et al., 2009; Huckstorf and Wirkner, 2011). The injection of resin liquid into the vascular system followed by resin curing and dissolution of the surrounding biological tissue gives a precise cast of the vascular structure. The obtained cast shows a structure with micrometer-scale fineness that can be visualized by x-ray microtomography.

*2.4. Resin embedding*

It takes from a few minutes to several tens of minutes to acquire x-ray absorption images for microtomographic visualization. It takes much more time to acquire phase contrast images. Sample drift or deformation faster than this data acquisition process causes artifacts in the resultant microtomogram. Therefore, soft tissue samples should be pretreated to fix all of the structural constituents for the data acquisition duration or to synchronize the structural change with the acquisition timing. Sample pretreatment for x-ray microtomography depends largely on sample source, state, size, and staining. Critical-point-dried samples can be visualized without sample rigidification. However, the problem most likely to occur in the data acquisition process of x-ray microtomography is sample drift. We recommend resin embedding when soft tissue samples are to be visualized at micrometer to submicrometer resolution by using synchrotron radiation microtomographs. Resin embedding rigidifies the sample structure at the nanometer scale and also facilitates the securing of the sample on the sample stage of microtomographs.

Epoxy resin is a popular material used in electron microscopy to rigidify biological tissue and enable it to be sliced with submicrometer thickness. Though most epoxy resins used for electron microscopy can also be used for x-ray microtomography, the transparency of the resin should be taken into account since sample position is usually confirmed with a light microscope while the sample is being mounted on the microtomograph. From this reason, we use a rather transparent resin designed for petrography (Petropoxy 154, Burnham Petrographics, ID) to embed biological samples for x-ray microtomography.

The epoxy resin should be degassed before use for several hours in vacuum to avoid bubbling in the subsequent curing process. The sample dehydrated in absolute ethanol is sequentially equilibrated with *n*-butylglycidyl ether and liquid resin. Since soft tissue samples equilibrated with liquid resin become transparent, the region of interest can be further dissected in this step. Tissue sample with a width less than 1 mm can be embedded in a borosilicate glass capillary (W. Müller Glas, Germany), as described previously (Mizutani et al., 2009b). Larger samples with widths of 2 mm to 5 mm can be embedded in resin pellets (Mizutani et al., 2008a) using silicone rubber tubing. Fig. 5 shows examples of resin-embedded samples.

*2.5. Sample mounting*

Data collection in x-ray microtomography is performed by rotating the sample while taking x-ray images. The sample should rotate exactly following the rotation of the sample stage. Several methods for mounting a biological sample on the microtomograph's sample stage have been reported (Metcher, 2009; Mizutani et al., 2009b).

In our work (e.g., Mizutani et al., 2007), since an x-ray image detector and other sample observation apparatus were placed in close proximity to the sample (Fig. 6a), we used brass fittings specially prepared for mounting capillary or resin pellet samples. The upper end of the fitting was designed according to the sample shape. Several examples of brass fittings are shown in Figs. 6b–d. The lower base of the fitting was secured on the sample stage with cap screws.

Capillary-embedded samples were mounted by inserting the capillary into a hole in the brass fitting filled with mounting clay specifically intended for crystallography (Fig. 6b). The diameter of the thin end of the brass fitting is 2.0 mm. This thin end allowed sample placement close to the detector face and minimized refraction effects in simple projection images. The lengths of the thin ends varied from 25 mm to 35 mm to bring the sample into the adjustable range of the stage height.

It is difficult to fix capillaries thinner than 0.5 mm using clay. Such capillaries were sleeved with brass tubing using epoxy glue. These samples were fixed by tightening the brass sleeve with a screw in a side face of the brass fitting (Fig. 6c). This fitting had a diameter of 10 mm and height of 17 mm.

Pellet samples prepared using silicone tubing were attached on the flat end of the brass fitting, as shown in Fig. 6d, using double-sided adhesive tape or epoxy glue. A sample fixed with glue could be easily removed from the fitting by snapping at the glued interface.

*2.6. Data collection*

*2.6.1 X-ray source*

X-ray sources used for microtomography are classified into two types: laboratory x-ray generator and synchrotron radiation source. Data acquisition is performed by placing samples in x-rays generated from these sources. Each voxel value of the resultant microtomogram represents the sample density observed with an x-ray spectrum defined by the source.

Most laboratory microtomographs are equipped with microfocus x-ray generators. Because the microfocus x-ray generator is usually used without a monochromator, its x-ray spectrum is a superposition of white x-rays with a broad profile and characteristic x-rays with sharp peaks. A sample image taken with the laboratory microtomograph, therefore, represents the overall transmittance of the white and characteristic x-rays. The high-energy component of the x-ray spectrum can readily pass through the sample, while the low-energy x-rays are absorbed by the sample. From this reason, voxel values of a 3D image taken using a laboratory microtomograph do not represent the LAC.

For synchrotron radiation microtomographs, radiation from the storage ring is usually monochromated to obtain x-rays of a certain energy. The available energy and brilliance of the monochromatic beam depend on the radiation source, monochromator, and other beamline components. The application of monochromatic x-rays to microtomography has several advantages. First, each voxel value of the 3D image taken using monochromatic x-ray simply corresponds to the LAC, which can be expressed as the weighted average of the LACs of atoms contained in that voxel. If the chemical components of the sample are determined in advance, each voxel is easily identified as one of the components (Tsuchiyama et al., 2005). Second, specific visualization of each probe element can be performed by using monochromatic x-rays with the energy corresponding to the absorption edge of the probe element (Mizutani et al., 2008c). Variability in the x-ray energy of synchrotron radiation allows us this specific visualization. Third, monochromicity is essential for the application of x-ray optical devices, such as a Fresnel zone plate, to microtomography. Submicrometer resolution has been achieved by using the zone plate optics at synchrotron radiation facilities (Winarski et al., 2010; Andrews et al., 2010; Uesugi et al., 2011). Besides these advantages, high-resolution microtomography relies on the highly brilliant and parallel characteristics of the synchrotron radiation. The parallelism of the synchrotron radiation is essential for taking x-ray images, especially in the simple projection geometry. X-ray brilliance that can effectively illuminate a small area within a short period of time is important for submicrometer resolution (Winarski et al., 2010; Andrews et al., 2010; Uesugi et al., 2011) and time-resolved analyses (Mokso et al., 2011).

*2.6.2 Spatial resolution and viewing field*

Resolution and viewing field are important concerns in structural analysis. In x-ray microtomography, these two parameters are defined by the x-ray optics and detector configurations. 3D visualization at nanometer-scale resolution has been achieved with synchrotron radiation microtomographs (Stampanoni et al., 2010; Winarski et al., 2010; Andrews et al., 2010; Uesugi et al., 2011) and laboratory microtomographs (Tkachuk et al., 2007; Van Loo et al., 2010; Sasov et al., 2011). X-ray diffraction microscopy of a single cell at nanometer resolution has also been reported (Nelson et al., 2010). However, structural analysis at a finer resolution naturally confines the viewing field to a smaller region. If the structure of a cellular network rather than a single cell is important for the sample's function, then the viewing field should contain a number of cells. Therefore, sample images are taken at a resolution appropriate for resolving the sample structure. The spatial resolution that should be achieved depends solely on the fineness of the structure to be visualized.

X-ray images are usually digitized into pixels for computational processing. The pixel dimension is typically set to half the x-ray optics resolution in order to minimize the amount of raw data. This Nyquist criterion has also been applied to image digitization in clinical CTs (Lin et al., 1993; Goldman, 2007). However, $N$-dimensional images should be taken with a pixel width of less than $1/2\sqrt{N}$ times the spatial resolution. Sampling at half the spatial resolution is insufficient for microtomographic visualization of 3D objects. The 3D image should be digitized with a pixel width of $1/2\sqrt{3}$ times the spatial resolution (Mizutani et al., 2010b, 2010c).

The numbers of pixels in the horizontal and vertical directions of x-ray images are governed by the specifications of the image detector. The product of the effective pixel size of the x-ray image and number of pixels along each direction gives the viewing field dimensions. The dimension of the viewing field along the sample rotation axis can be extended by simply shifting the sample along the rotation axis and stacking the obtained microtomograms (Fig. 7a). The dimensions perpendicular to the axis can be extended by taking 3D images of the sample in a honeycomb lattice manner (Fig. 7a), though forming several datasets into the entire structure is more complex than stacking. Instead, the dimensions perpendicular to the rotation axis are doubled by offset microtomography (Mizutani et al., 2010d). In this offset microtomography, x-ray images in the range from 0° to 360° instead of from 0° to 180° in usual microtomography are taken by placing the rotation axis near the left or right end of the detector's viewing field (Fig. 7a).

*2.7. Microtomographic reconstruction*

In x-ray microtomography, sequential two-dimensional (2D) images are acquired while the sample is being rotated (Fig. 7b). The horizontal pixel strip extracted from these 2D images is subjected to Fourier transformation, the application of a filter function, and inverse Fourier transformation. Then back-projection calculations are performed to obtain the microtomogram. The overall strategy of this microtomographic reconstruction calculation is the same as for clinical CT.

Since Fourier transformation and back-projection calculations are iterations of simple arithmetic, microtomographic reconstruction can be accelerated by using a parallel computing environment. For reconstruction calculation using PCs, the CUDA or OpenCL environment commercially available as a graphics card is appropriate and cost efficient. A reconstruction program that we have developed (http://www.el.u-tokai.ac.jp/ryuta/) works for both of these environments.

Since each microtomographic slice plane perpendicular to the sample rotation axis is reconstructed through the abovementioned calculation, the in-plane resolution within the reconstructed microtomographic slice is affected by the reconstruction calculation while the through-plane resolution along the rotation axis is mainly independent of the reconstruction calculation. These in-plane and through-plane resolutions should be differentiated, especially if the 2D images are taken with a pixel width coarser than $1/2\sqrt{3}$ times the spatial resolution. The spatial resolution of resultant microtomograms can be examined from the intensity profile of periodic patterns (Mizutani et al., 2008d, 2010b) or sample surfaces (Seifert and Flynn, 2002; Mizutani et al., 2010c).

The process that follows the reconstruction calculation differs from sample to sample. Least-squares fitting of 3D images should be performed to obtain a differential image at the x-ray absorption edge of the probe element. Least-squares fitting is also performed to compose microtomograms from multiple datasets taken by shifting the sample. In the case of capillary-embedded samples, the capillary image surrounding the sample interferes with the evaluation and analysis of the sample's 3D structure. Therefore, the capillary should be striped from the 3D image by replacing voxels that correspond to the capillary wall with null coefficients. This can be done by determining the capillary surface by using a certain LAC threshold and erasing voxels to a predefined depth from that surface.

**3. Microtomography in biological applications**

Applications of x-ray microtomography in biology have been widely reported for a great number of objects. The leading application of x-ray microtomography in biology is the structural study of bones and teeth. Microtomographic analyses of soft tissues are rather difficult but can reveal the functional mechanisms of 3D cellular and subcellular structures. Here, we look at several examples of the application of x-ray microtomography to soft tissues.

*3.1. Neuronal network*

Neuronal circuits, which are essential for brain functions, are built up by neurons as a 3D network, so tracing the 3D neuronal network of human cerebral cortex is the first step to understanding the mechanism of human brain functions. We have demonstrated that a skeletonized model of human neurons can be built by tracing the 3D coefficient map obtained by x-ray microtomography (Mizutani et al., 2010a). In this process, the 3D image was converted into 3D Cartesian coordinates by model building (Mizutani et al., 2011). The 3D coordinates are easier to handle than the image itself, making it possible to analyze human brain circuits (Fig. 8). Such analysis has revealed that a flip-flop local circuit is one of the units composing our brain network (Mizutani et al., 2010a). This analysis can be performed for any part of the brain. Analysis of the circuits of the whole brain in the future will enable the simulation of human brain functions including thought, imagination, and self-consciousness. The mechanisms of these brain functions can be unveiled by deciphering the machine language code executed in human brain circuits.

*3.2. Morphology*

The morphology of an organ is important for understanding its function. As has been discussed, x-ray microtomography can be used for exact volumetric measurements and the determination of mechanical properties of myocardium (Happel et al., 2010). The differentiation of fine structural and morphological details such as the abdomen with the intestine and the genital organs of an insect has been reported (Zhang et al., 2010). The obtained structure can provide a structural insight into biological function. The microtomographic structure of coxa-trochanteral joints of the legs of a Papuan weevil exhibited a functional screw-and-nut system found for the first time in biological organisms. (van de Kamp et al., 2011). It has been pointed out that the non-destructive nature of x-ray microtomography is an important consideration when dealing with scarce materials such as museum specimens of extinct or rare species that are essential for resolving evolutionary trends (Jeffery et al., 2011).

There have also been applications of biological x-ray microtomography in food science. Fat content in a meat sample can be three-dimensionally visualized (Frisullo et al., 2010), indicating that x-ray microtomography is a useful technique for non-invasive visualization and measurement of the internal microstructure of cellular food products.

*3.3. Tissue engineering*

Recent progress in stem cell technology (Yamanaka and Blau, 2010) has provided a clue to the regeneration of damaged tissues. The next step for tissue regeneration is to build the 3D architecture of the target tissue, since the tissue's 3D structure is responsible for its biological functions. X-ray microtomography has significant potential in the characterization of 3D scaffolds and tissue-engineered bones (Ho and Hutmacher, 2006; Cancedda et al., 2007; Faraj et al., 2009). We have reported that the microtomographic structure of soft tissues can be used as a 3D template for the artificial fabrication of biological architectures (Mizutani et al., 2008b). The 3D data necessary to create scaffolding that directly mimics the structural patterns of a microvascular tree system has been determined by x-ray microtomography (Mondy et al., 2009). X-ray microtomography using a contrast agent can be applied to the nondestructive 3D assessment of neovascularization within a healing bone defect (Young et al., 2008). The microtomographic visualization of tissues and organs should provide the 3D microstructural basis that allows the artificial regeneration of biological tissues.

**4. Perspectives**

Visualization of 3D structures at submicrometer resolution has been achieved through progress in x-ray microscopy and x-ray optical devices (Stampanoni et al., 2010; Winarski et al., 2010; Andrews et al., 2010; Uesugi et al., 2011). Sub-second temporal resolution has also been achieved and this makes it possible to visualize dynamic processes (Mokso et al., 2011). Efforts to improve the spatial and temporal resolutions will lead to the visualization of 3D microstructures that have not been determined.

Recent studies on x-ray fluorescence microtomography allowed observation of the 3D elemental distribution. A 3D visualization of calcium localization at submicrometer resolution has been performed by using the combination of a third-generation x-ray undulator and zone plate optics. (Watanabe et al., 2009). The use of a monochromatic x-ray beam focused to a diameter of 1 μm to 10 μm revealed multi-element 3D distributions at concentrations down to 1 µg/g (Lanzirotti et al., 2010). X-ray fluorescence microtomography at 400 nm resolution has been applied to subcellular imaging of the whole cell of a diatom, quantifying the 3D distributions of several elements (de Jonge et al., 2010). These microtomographic studies were performed by scanning a sample with a thin x-ray beam and they needed much more data acquisition time than usual x-ray microtomography. Since x-ray fluorescence microtomography can resolve the 3D distribution of elements at concentrations of the order of 1 ppm, its biological applications will increase if the fluorescence dataset can be acquired in a shorter time.

Although microtomographic studies have been reported for a number of biological objects, the scientific understanding of 3D microtomographic images is still in progress. This is mainly due to the complexity of the 3D structures of biological objects. Simplification of the 3D microtomographic image should facilitate understanding of complicated microstructures composed of huge numbers of voxels. Structural analyses of tomographic images based on model building have been reported for a number of organs, e.g., pulmonary airway (Chaturvedi and Lee, 2005) and intervertebral disc (Natarajan et al., 2004). Microtomographic visualization followed by 3D analysis allows us to obtain quantitative results from 3D images (Brabant et al., 2011). We have recently reported microtomographic analysis of human brain circuits (Mizutani et al., 2010a). The analysis of the neuronal circuits has been achieved by building skeletonized models of neurons in the 3D coefficient map determined by x-ray microtomography (Mizutani et al., 2011). Since the 3D structure of brain tissue is highly complicated, the neuronal circuits cannot be resolved just by viewing the 3D structure. Analytical resolution of the neuronal circuits has been attained for the first time after tracing the 3D coefficient distribution.

Biological tissues exhibit complicated structures from the organ level to the subcellular level. Model building in the microtomographic image at cellular to subcellular resolution can reveal the structural basis of biological function embedded in the 3D image. A methodology for building simplified models in the 3D coefficient map determined by x-ray microtomography should be established and refined in order to unveil microstructures responsible for biological function.

It has been difficult to determine real 3D structures of cellular and subcellular constituents of biological soft tissues. Hence, most studies in the structural biology discipline have been focused on molecular structures rather than on higher-order structures composed of cells or subcellular organelles. Recent application of x-ray microtomography in biology has revealed such cellular and subcellular structures relevant to biological functions. The convergence of sciences in x-ray optics, biology, and 3D structure analysis in biological microtomography should break new ground with the interdisciplinary approach. We suggest that x-ray microtomography will become increasingly important as a method for unveiling 3D structures in biology, like x-ray crystallography in molecular biology.

**Acknowledgments**

We thank Yasuo Miyamoto, Technical Service Coordination Office, Tokai University, for helpful assistance with microfabrication. We thank Kiyoshi Hiraga, Technical Service Coordination Office, Tokai University, for preparing brass fittings. We thank Hiroki Shimamura, School of Engineering, Tokai University, for assistance with sample preparation. This work was supported in part by Grants-in-Aid for Scientific Research from the Japan Society for the Promotion of Science (no. 21611009). The synchrotron radiation experiments were performed at SPring-8 with the approval of the Japan Synchrotron Radiation Research Institute (JASRI) (proposal nos. 2011A0034, 2011B0034, and 2011B0041).

**Web references**

**Figure captions**

**Fig. 1.** Structures of white matter of human frontal cortex visualized by osmification. Microtomographic data collection was performed at BL20XU of SPring-8 using 12-keV x-rays. LACs are shown in gray scale from 25.2 cm$^{-1}$ (black) to 34.0 cm$^{-1}$ (white). Scale bar: 50 µm. Reproduced from Mizutani et al. (2009b).

**Fig. 2.** Stereo drawing of intestinal villi of adult C57BL/6 mouse visualized by the periodic acid-methenamine-silver stain method. Microtomographic data collection was performed at BL20XU of SPring-8 using 12-keV x-rays. LACs are shown in gray scale from 10 cm$^{-1}$ (black) to 50 cm$^{-1}$ (white). Scale bar: 50 µm.

**Fig. 3.** Stereo drawing of neurons in human frontal cortex visualized by Golgi silver impregnation. Microtomographic data collection was performed at BL20XU of SPring-8 using 12-keV x-rays. LACs are shown in gray scale from 10 cm$^{-1}$ (black) to 50 cm$^{-1}$ (white). Scale bar: 50 µm. Reproduced from Mizutani et al. (2008a).

**Fig. 4.** Example of absorption-contrast microtomogram of epoxy resin filling a borosilicate glass capillary. Microtomographic data collection was performed at BL20XU of SPring-8 using

12-keV x-rays. Total duration of x-ray exposure was 216 s. LACs are shown in gray scale from -10 cm$^{-1}$ (black) to 10 cm$^{-1}$ (white). Box size: 100 μm × 100 μm.

**Fig. 5.** Resin-embedded samples. *A*, pellet sample cured using silicone rubber tubing and a polypropylene case. *B*, pellet taken out of the silicone tubing. *C*, sample embedded in a borosilicate glass capillary.

**Fig. 6.** (a) Capillary-embedded sample mounted using a brass fitting. The brass fitting was attached to the x-ray microtomograph installed at BL20XU of SPring-8. The x-ray image detector was placed close to the sample to minimize refraction effects. (b) Capillary-embedded sample inserted in a 1.0- or 1.5-mmϕ hole of a brass fitting filled with clay. The diameter of the thin end of the brass fitting is 2.0 mm. The lengths of the thin ends vary from 25 mm to 35 mm. (c) Capillaries thinner than 0.5 mm were sleeved with brass tubing and fixed with a screw in a side face of the brass fitting. The fitting had a diameter of 10 mm and height of 17 mm. Reproduced from Mizutani et al. (2009b). (d) Pellet sample attached on the flat end of a brass fitting using double-sided adhesive tape. The upper end of the brass fitting vary from 3.0 to 5.0 mm in diameter and 15 to 30 mm in height.

**Fig. 7.** (a) Viewing field in microtomography. Cylinders indicate viewing field volume. Broken lines indicate sample rotation axis. *A*, Data acquisition with the usual microtomography is performed by taking 2D images indicated with the hatched box in the range from 0° to 180°. The resultant viewing field can be represented as a cylinder. *B*, In offset microtomography, 2D images in the range from 0° to 360° are taken by placing the rotation axis near the left or right end of the viewing field. *C*, The dimension of the viewing field along the sample rotation axis can be extended by simply shifting the sample along the rotation axis and stacking the obtained 3D images. *D*, Dimensions perpendicular to the rotation axis can be extended by taking 3D images in a honeycomb lattice manner. (b) Principle of the microtomographic reconstruction. *A*, 2D images are taken by rotating the sample. *B*, The horizontal pixel strip extracted from the 2D image is subjected to Fourier transformation, application of a filter function, and inverse Fourier transformation. *C*, Back-projection calculations are then performed to obtain the microtomogram.

**Fig. 8.** Schematic diagram of microtomographic analysis of brain circuits. Microstructure of human brain tissue was determined by x-ray microtomography (*A*). The obtained 3D distribution of absorption coefficient were used for tracing (*B*), giving neuronal networks (*C*). Neuronal circuits were analytically resolved from the traced networks (*D*). Reproduced from

Mizutani (2010).

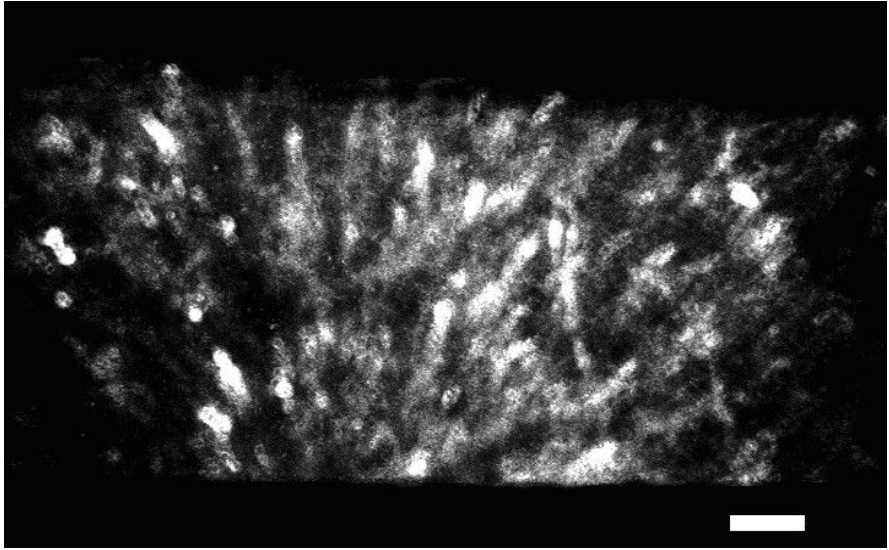

Fig. 1.

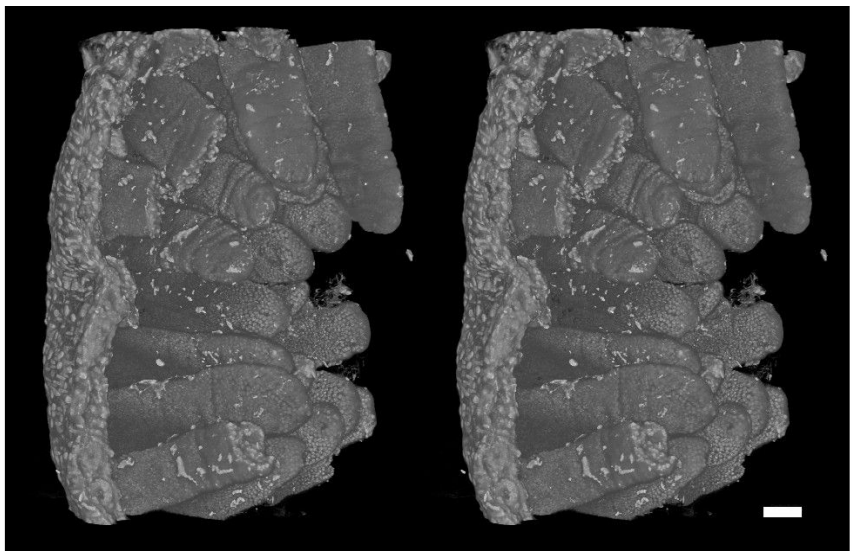

Fig. 2.

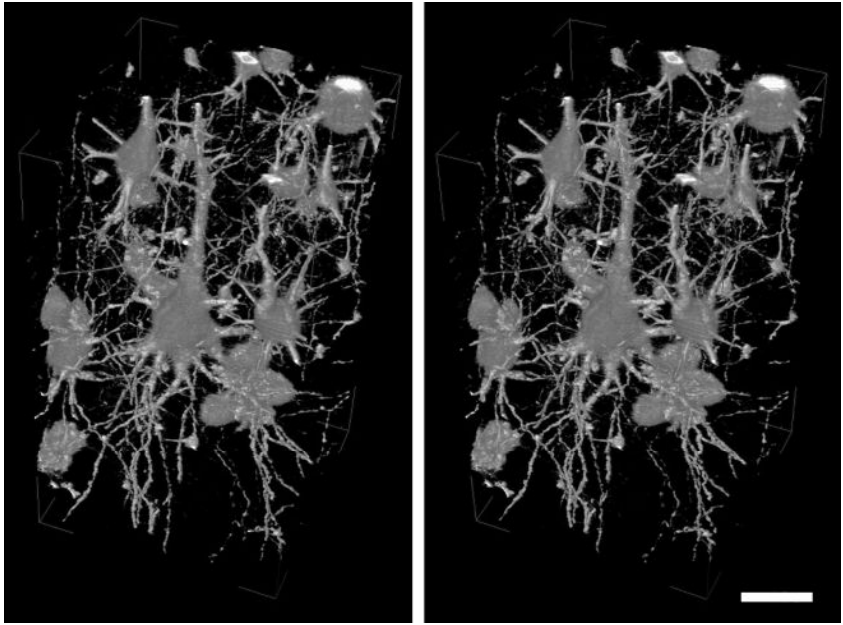

Fig. 3.

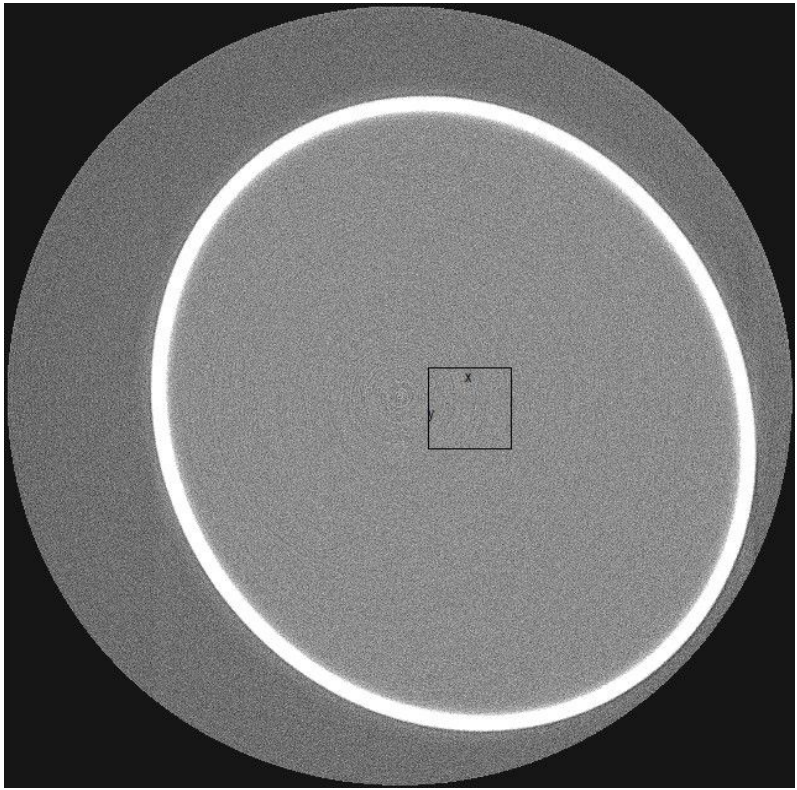

Fig. 4.

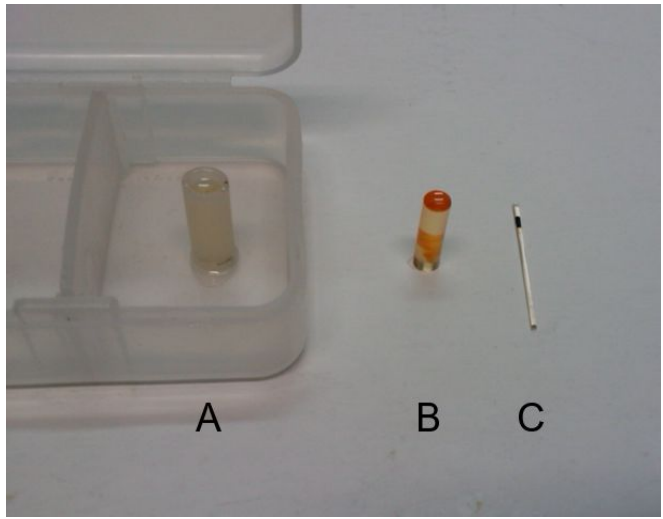

Fig. 5

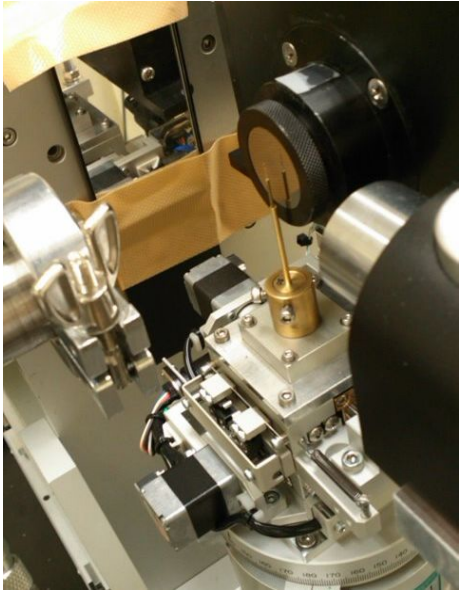

Fig. 6(a)

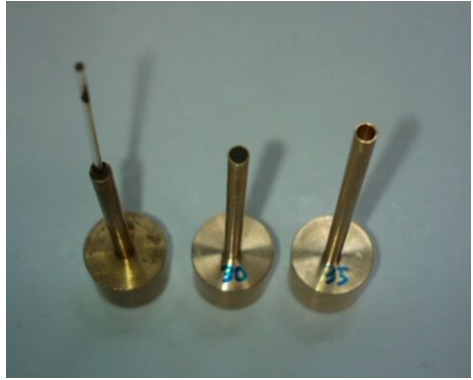

Fig. 6(b)

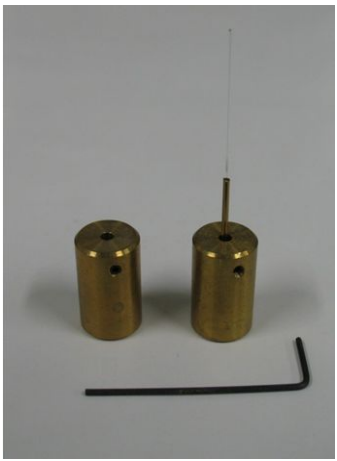

Fig. 6(c)

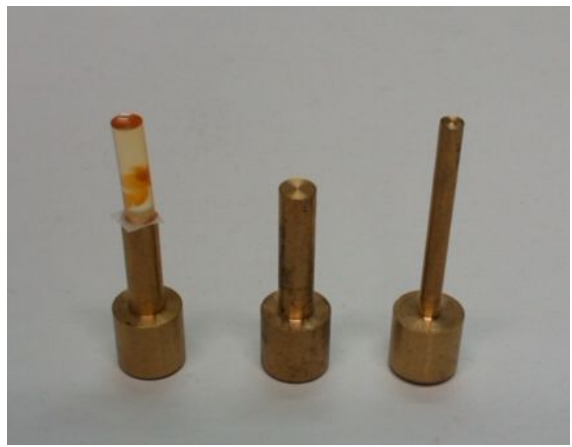

Fig. 6(d)

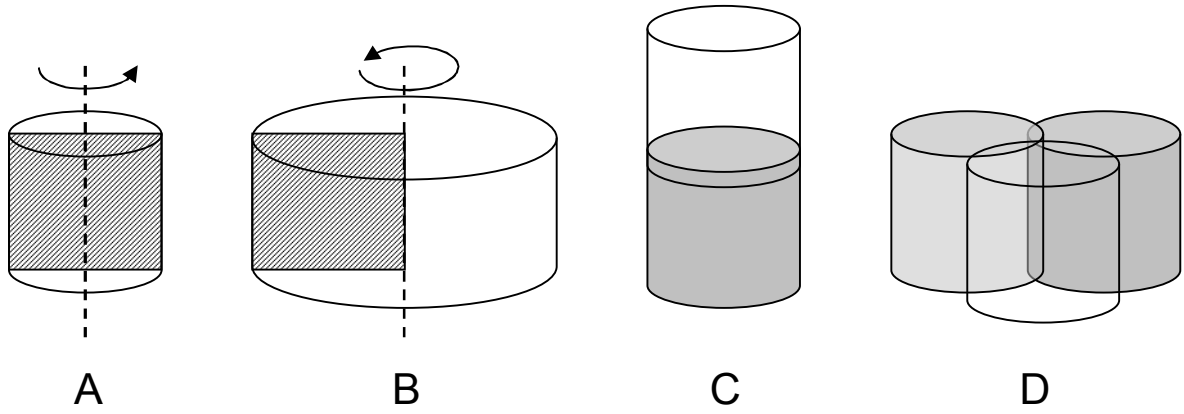

Fig. 7(a)

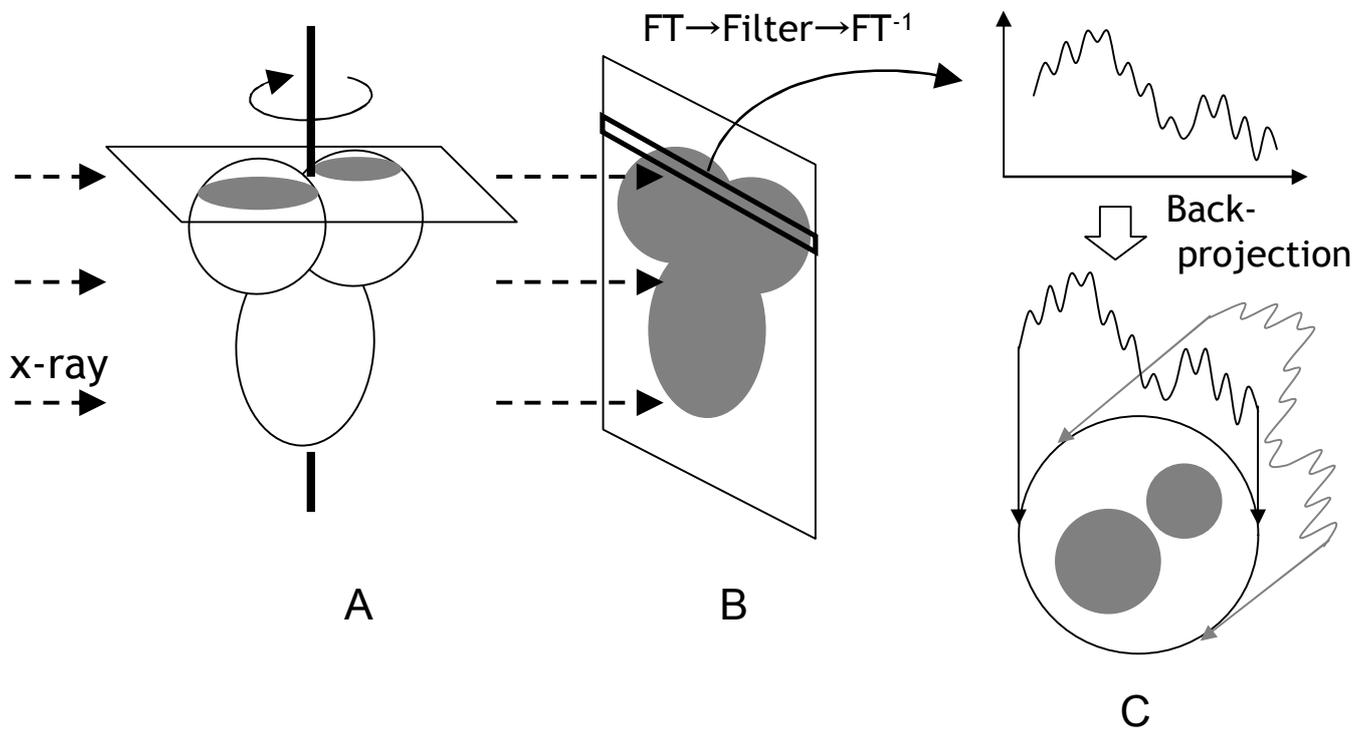

Fig. 7(b)

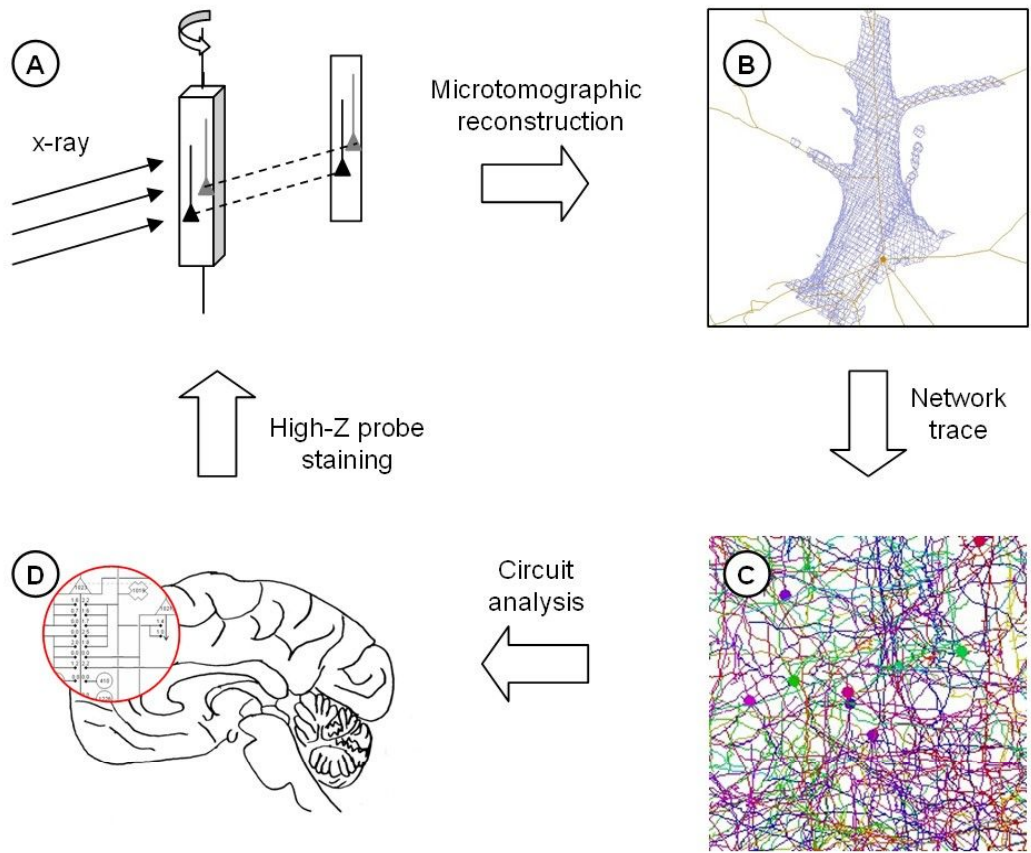

Fig. 8